\documentclass[conference]{IEEEtran}
\IEEEoverridecommandlockouts
\usepackage{cite}
\usepackage{amsmath,amssymb,amsfonts}
\usepackage{algorithmic}
\usepackage{graphicx}
\usepackage{textcomp}
\def\BibTeX{{\rm B\kern-.05em{\sc i\kern-.025em b}\kern-.08em
    T\kern-.1667em\lower.7ex\hbox{E}\kern-.125emX}}

\usepackage[table,x11names]{xcolor}
\usepackage{xfrac}
\usepackage{pgfplots}
\usepackage{tikz}
\usepackage{amsfonts}
\usepackage{amssymb}
\usepackage{amsbsy}
\usepackage{amsthm, mathrsfs}
\usepackage{mathtools}
\usepackage{array} 
\usepackage{cite}
\usepackage{dsfont}
\usepackage{graphicx}
\usepackage{subcaption}
\usepackage{caption}
\usepackage{tabularx}

\usepackage{wrapfig}

\usepackage{multirow}
\usepackage{rotating}

\theoremstyle{definition}

\usepackage{paralist}
 
  \usepackage{url}

\usepackage{pgfplots}

\usepackage{xfrac}

\usepackage{accents}

\begin{document}

\title{Privacy-Preserving Identification via\\ Layered Sparse Code Design: Distributed Servers and Multiple Access Authorization\\
\thanks{This research has been supported by the ERA-Net project ID\_IoT No 20CH21\_167534.}
}

\author{\IEEEauthorblockN{Behrooz~Razeghi, Slava~Voloshynovskiy, Sohrab Ferdowsi and Dimche Kostadinov}
\IEEEauthorblockA{Stochastic Information Processing Group\\ Department of Computer Science, University of Geneva, Switzerland\\
 {\small \{\texttt{behrooz.razeghi, svolos, sohrab.ferdowsi, dimche.kostadinov}\}@\texttt{unige.ch}}}}
 

\maketitle

\begin{abstract}
We propose a new computationally efficient privacy-preserving identification framework based on layered sparse coding. 
The key idea of the proposed framework is a sparsifying transform learning with ambiguization, which consists of a trained linear map, a component-wise nonlinearity and a privacy amplification. We introduce a practical identification framework, which consists of two phases: public and private identification. The public untrusted server provides the fast search service based on the sparse privacy protected codebook stored at its side. The private trusted server or the local client application performs the refined accurate similarity search using the results of the public search and the layered sparse codebooks stored at its side. The private search is performed in the decoded domain and also the accuracy of private search is chosen based on the authorization level of the client. 
The efficiency of the proposed method is in \textit{computational complexity} of encoding, decoding, ``encryption" (ambiguization) and ``decryption'' (purification) as well as \textit{storage complexity} of the codebooks. 
\end{abstract}


%
\IEEEpeerreviewmaketitle

 
 \vspace{-6pt}

\section{Introduction}

\vspace{-4pt}

Privacy-preserving identification is of great importance for the growing amount of applications that require fast and accurate identification. Third parties are assumed to perform the expected services but are curious about the nature of the data content of the queries. These applications include but are not limited to the internet-of-things (IoT), biometrics, clinical reports, etc.

\vspace{-1pt}

In this work, we propose a new distributed framework of privacy-preserving identification based on successive refinement. 
The successive refinement of information was first studied for the classic source coding problem \cite{equitz1991successive}. 
The performance of this problem is formulated by a rate-distortion theory. The objective is to achieve the rate-distortion bound at each stage. 
In \cite{Sohrab_WIFS2016} the authors proposed the Sparse Ternary Coding (STC) scheme for fast search in large scale identification problems. The theoretical results of the STC scheme are studied in \cite{Sohrab_ISIT2017}. Inspired by the successive refinement of information problem, the authors proposed a multi-layer network which successively generates sparse ternary codes, which closely achieve the Shannon lower bound of the distortion-rate function. 


\vspace{-7pt}
 
\subsection{Our Contribution} 
 
 \vspace{-5pt}

In this paper, we propose a new framework of multi-stage identification using successive refinement with sparse ternary codes at each layer of the privacy-preserving identification. The proposed privacy-preserving mechanism is based on the ambiguization, i.e., addition of noise to non-zero sparse data representation in the transform domain. We demonstrate that the security of this scheme does not rely on the secrecy of transform. Accordingly, we develop a distributed search framework (Fig. 1) with a granted granular access to the results of the search based on the level of authorization expressed in the knowledge of codebook and vote refinement. We demonstrate that the identification based on compressed STC representation could be a good first stage for the fast public identification, while the authorized private users can benefit from the refined results enjoying the accurate upgrades in the reconstructed real space with a low computational complexity. Up to our best knowledge, the proposed scheme is among the first that is based on the successive refinement with the sparse ternary coding bridging the gap to the theoretical performance limits.

%

 \begin{figure}[!t]
\centering
\includegraphics[scale=0.5]{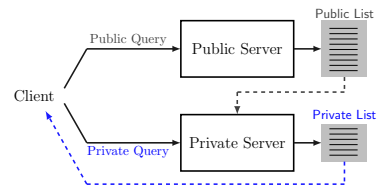}
\caption{General block diagram of the proposed framework.}
\vspace{-17pt}
 \label{Fig:GeneralBlockDiagram}
\end{figure} 

\vspace{-7pt}

 \subsection{Notation}

 \vspace{-4pt}
 
Matrices and vectors are denoted by boldface upper-case ($\mathbf{X}$) and lower-case ($\mathbf{x}$) letters, respectively. 
We consider the same notation for a random vector $\mathbf{x}$ and its realization. The difference should be clear from the context. $x_i$ denotes the $i$-th entry of vector $\mathbf{x}$. For a matrix $\mathbf{X}$, 
$\mathbf{x}{\left(  j \right)}$ denotes the $j$-th column of $\mathbf{X}$. The superscript $(\cdot)^\dagger$ stands for the pseudo-inverse and $(\cdot)^T$ stands for the transpose. We use the notation $\left[ N \right]$ for the set $\{ 1, 2, ..., N\}$.
 
 \vspace{-7pt}

\subsection{Outline of the Paper}

\vspace{-4pt}

The remainder of the paper is organized as follows. In Sec. \ref{Sec:II}, the problem formulation is introduced. Then, in Sec. \ref{Sec:III} we present our framework. We provide the performance analysis in Sec. \ref{Sec:IV}. Finally, conclusions are drawn in Sec. \ref{Sec:V}.

\vspace{-7pt}

\section{Problem Formulation}\label{Sec:II}

 \vspace{-3pt}

Suppose that an owner has a collection of $M$ raw vectors $\mathbf{x}{\left(m\right)}, m \in \left[M\right]$ in the database $\mathbf{X} =  \left[ 
\mathbf{x}{\left(1\right)},   \cdots ,\mathbf{x}{\left(M\right)}
\right]$, 
where each raw vector $\mathbf{x}{\left(m\right)}, m \in \left[M\right]$ from a set $\mathcal{X} \subset {\mathbb{R}}^{N}$ is a random vector with distribution $p\left( \mathbf{x} \right)$ and bounded variance $\sigma^2_{\mathbf{x}}$.
In general, the input data might be raw or based on extracted features such as those from (aggregated) local descriptors \cite{jegou2009burstiness, perronnin2007fisher, jegou2010aggregating}, or the top layer of a neural network \cite{babenko2014neural} or the latent space of auto-encoders \cite{kingma2014auto}. 
%
The user has a query $\mathbf{y}{\left(m\right)}  \in {\mathbb{R}}^N$ which is a noisy version of $\mathbf{x}{\left(m\right)}$, i.e., $\mathbf{y}{\left(m\right)} = \mathbf{x}{\left(m\right)} + \mathbf{z}$, where we assume $\mathbf{z} \in {\mathbb{R}}^N$ is a Gaussian noise vector with distribution $\mathcal{N} \! \left( \mathbf{0},   \sigma^2_{\mathbf{z}} \mathbf{I}_N \right)$. The user is interested in some information about the subset $\mathcal{L} \left( \mathbf{y} \right)$ of the $\gamma$-NN (or $\gamma$-ANN) of $\mathbf{y}$. The owner subcontracts the similarity search to an entity called the server. 
The clients and data owner attempt at protecting their data from (public) server side analysis, which is assumed to be honest but curious.

\begin{figure}[!t]
\centering
\includegraphics[scale=0.46]{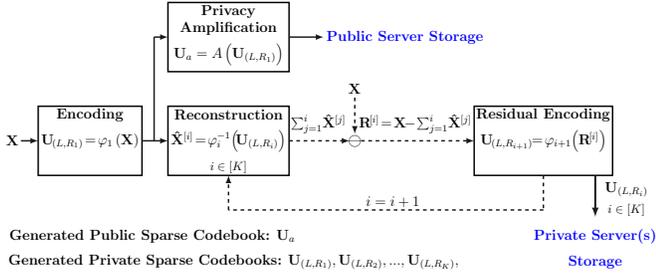}
\caption{Successive sparse codebooks generation scheme.}
\vspace{-15pt}
 \label{Fig:CodebookGeneration}
\end{figure}

 \vspace{-5pt}

\section{Proposed Framework}\label{Sec:III}
 
  \vspace{-3pt}
 
  \subsection{Framework Overview}
  \vspace{-3pt}
  
Our framework is composed of the following five steps:
 
 \vspace{-2pt}
\subsubsection{Preparation at Owner Side}

The owner generates one \textit{public sparse codebook} plus $K$ \textit{private sparse codebooks} from the media data that he owns (Fig.~\ref{Fig:CodebookGeneration}). The public codebook is sent to the public storage server (e.g., Google Site) and the $K$ private sparse codebooks are sent to the private server storage (e.g., ``Friend" Sites). The public sparse codebook is generated using the \textit{learned sparsifying transform} followed by an \textit{element-wise nonlinearity} and a \textit{privacy amplification}. The $K$ private codebooks are generated by the successive refinement encoder that will be explained in the text below.  
 
 
  \vspace{-2pt}
\subsubsection{Indexing at Server Sides}
The public and private servers index the received sparse codes. 

\vspace{-2pt}
 
\subsubsection{Querying at Client Side}

The client generates a sparse code from his query data using the same transformation scheme used for \textit{public} search (Fig.~\ref{Fig:PublicIdentification}). 
Then, the client sends the sparse code of his query data to the public server and his original domain query to the private server.


 \vspace{-2pt}

\subsubsection{Initial Search at Public Server Side} 

\begin{figure}[!t]
\centering
\includegraphics[scale=0.47]{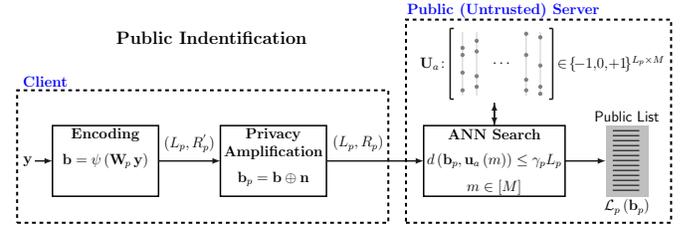}
\caption{Public identification scheme.}
\vspace{-15pt}
 \label{Fig:PublicIdentification}
\end{figure}

The server runs a similarity search to identify the sparse codes that are most similar to the query (Fig.~\ref{Fig:PublicIdentification}). The public list, which consists of indices of the most similar codes, is sent to the private server. 


 \vspace{-2pt}

\subsubsection{Multi-layer List Refinement at Private Server}

\begin{figure}[!t]
\centering
\includegraphics[scale=0.445]{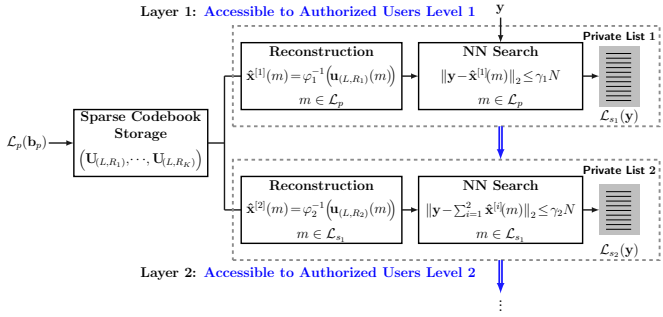}
\caption{Private multiple access identification scheme.}
\vspace{-7pt}
 \label{Fig:PrivateIdentification}
\end{figure}

The private server looks at his first layer codebook and decodes (reconstructs) the sparse codes that are within the public list. 
Then he runs a similarity search using the received query and the decoded sparse codes, i.e., the similarity is computed in the original domain. 
This similarity search results in the first private list, which is accessible to the authorized users at level 1. 
Next, the private server uses his second layer codebook and decodes the sparse codes with indices within the initial private list. The second private list is hereby computed using similarity search between the received query and \textit{superposition} of the decoded sparse codes of this layer and the previous layer. This list is accessible to the authorized users at level 2. Analogously, the private lists are refined successively by running the similarity search between the query and the superposition of the decoded sparse codes of each layer and all previous layers (Fig.~\ref{Fig:PrivateIdentification}).

%

 \vspace{-5pt}
 
\subsection{Layered Sparse Coding}


\subsubsection{Principal Element}

\vspace{-3pt}

The core of our coding paradigm is as follows:




\vspace{-2pt}

\textit{Encoder:} 
This is defined by a mapping $\varphi \! :  \!\! \mathbb{R}^N  \! \rightarrow \! \{ -1, 0 , +1 \}^L$. Given the (raw) feature vector $\mathbf{x} (m), m \! \in  \! [M]$ the encoder generates the sparse code $\mathbf{u}_{(L, R)}(m), m \!\in \![M]$ with dimensionality $L$ and rate $R$, therefore $\mathbf{u}_{(L,R)}(m) \! =\!  \varphi \left( \mathbf{x} (m)  \right)$. 


Indeed, our encoder is based on the sparsifying transform learning model \cite{ravishankar2013learning} followed by a non-linearity thresholding function to constraint the alphabet of codes. This model suggests that a feature vector $\mathbf{x}{\left(m\right)} \in \mathbb{R}^N$ is approximately sparsifiable using a transform $\mathbf{W} \in \mathbb{R}^{L \times N}$, that is $\mathbf{W} \mathbf{x}{\left(m\right)} = \mathbf{a}{\left(m\right)} + \mathbf{e}_{\mathbf{a}}$, where $\mathbf{a}{\left(m\right)} \in \mathbb{R}^L$ is sparse, i.e., ${\| \mathbf{a}{\left(m\right)} \|}_{0} \ll L$, and $\mathbf{e}_{\mathbf{a}} \in \mathbb{R}^L$ is the representation error of the feature vector or residual in the \textit{transform domain}. The sparse coding problem for this model is a direct constraint projection problem. This sparse approximation is as follows: \vspace{-4pt}
\begin{equation}\label{l_0-optimalsolution} 
\mathbf{\widehat{a}}{(m)} \!\!  =\!  \mathop{\arg \min}_{\mathbf{a}{\left(m\right)} \in \mathbb{R}^L} { \| \mathbf{W} \mathbf{x}{\left(m\right)} \!  -\!  \mathbf{a}{\left(m\right)} \|}_2^2 + \lambda \Omega \left(\mathbf{a}{\left(m\right)} \right)\! , \forall m \in \! \left[ M \right]. \nonumber \vspace{-4pt}
\end{equation} 

\vspace{-2pt}

The above direct problem has a closed-form solution for any of the two important regularizers $\Omega \left( \cdot \right) =  {\| \cdot \|}_0$ or $\Omega \left( \cdot \right) =  {\| \cdot \|}_1$. Analogous to \cite{Razeghi2017wifs}, we consider the $\ell_0$-``norm'' as our sparsity-inducing penalty. In this case, the solution $\mathbf{\widehat{a}}{(m)}$ is obtained exactly by hard-thresholding the projection $\mathbf{W} \mathbf{x}{\left(m\right)}$ and keeping the $S_x$ entries of the largest magnitudes while setting the remaining low magnitude entries to zero. For this purpose, we define an intermediate vector ${\mathbf{f}}{\left(m\right)}  \triangleq \mathbf{W} \mathbf{x}{\left(m\right)} \in \mathbb{R}^L$ and denote by $\lambda_x$ the $S_x$-th largest magnitude amongst the set $\{ \vert f_1{\left(m\right)}  \vert , ..., \vert f_L{\left(m\right)}  \vert  \}$. Then the closed-form solution is achieved by applying a hard-thresholding operator to $\mathbf{f}{\left(m\right)} $, which is defined as 
$\mathbf{a} \left(m\right) = H_{\lambda_x} \! \left( \mathbf{f} \left(m\right)  \right) \! = \mathds{1}_{\vert f_l {\left(m\right)}  \vert \ge \lambda_x } {\mathbf{f}}{\left(m\right)}  ,   \forall m \in \left[ M \right], \forall l \in \left[ L\right]$.
%
Now we impose extra constraint on the alphabet of our codes by applying the ternary hash mapping to $H_{\! \lambda_x} \! \!\left( \! \mathbf{W} \mathbf{x} (m)\!  \right)$ as: \vspace{-1pt}
\begin{equation}\label{Eq:OurEncoding}
\mathbf{u} \left(m\right) \triangleq \psi_{\lambda_x} \left(  \mathbf{W} \mathbf{x} \left(m\right)  \right) \in  {\{ -1, 0, +1 \}}^L , \;  \forall m \in \left[ M \right],
\vspace{-2pt}
\end{equation}
where 
$\psi_{\lambda_x} \left(  \mathbf{W} \mathbf{x} \left(m\right)  \right) =  \mathrm{sign}  \left( H_{\lambda_x} \left( \mathbf{W} \mathbf{x} \left(m\right) \right) \right) $. 
The bit rate of this code can be formulated as $R \! = \! \frac{1}{L} \! \log_2 \left( \!  \! \binom{L} {S_x} 2^{S_x} \! \right)$. 
We denote by $\varphi \left( \cdot \right)$ the encoder in general, therefore, the codeword $\mathbf{u} (m)$ with block-length $L$ and rate $R$ is denoted by $\mathbf{u}_{(L, R)} (m) = \varphi \left( \mathbf{x}(m) \right), m \in [M]$. 

In general, we have a joint learning problem that can be formulated as: \vspace{-1pt}
\begin{equation}\label{Eq:JointLearning}
\left( \! \mathbf{\hat{W}}, \! \mathbf{\hat{A}} \! \right) \! = \! \mathrm{arg} \!  \mathop{\min}_{\left( \mathbf{W}, \mathbf{A} \right)} \!  {\| \mathbf{W} \mathbf{X} \!  - \!  \mathbf{A} \|}_F^2 + \beta_W \Omega_W \!  \left( \mathbf{W} \right) + \beta_A \Omega_A \! \left( \mathbf{A} \right),
\end{equation}
where $\beta_W \geq 0$ and $\beta_A \geq 0$ are regularization parameters, $\Omega_W \!  \left( \mathbf{W} \right)$ and $\Omega_A \! \left( \mathbf{A} \right)$ are the constraints on the linear mapper $\mathbf{W}$ and sparse (but not ternarized) code matrix $\mathbf{A}$, respectively. 
The algorithm for the above problem alternates between solving for $\mathbf{A} = H_{\lambda} \left( \mathbf{W} \mathbf{X}\right)$ (sparse coding step) and $\mathbf{W} = \mathbf{U}_W \mathbf{V}_W^T$ (transform update step), whilst the other variables are kept fixed. 
Finally, the ternarized sparse codebook $\mathbf{U}_{(L, R)}$ is obtained as $\mathbf{U}_{(L, R)} = \varphi \left( \mathbf{X} \right)$, which consists of $M$ sparse codewords $\mathbf{u}_{(L, R)} (m) \in [M]$. 

\textit{Decoder:}  
This is a mapping $\varphi^{-1} \!: \! \{ -1, 0 , +1 \}^L \! \!\rightarrow \!\mathbb{R}^N$. Base on $\mathbf{u}_{(L, R)}(m)$ generated at the encoder, the decoder produces reconstruction $\mathbf{\widehat{x}} (m)= \varphi^{-1} \left( \mathbf{u}_{(L,R)}(m) \right) = \mathbf{W}^{\dag} \mathbf{u}_{(L,R)}(m)$. That is, our decoding is simply a pseudo-inverse operation.

\subsubsection{Overall Scheme}

In \cite{Razeghi2018icassp}, the authors studied the reconstruction performance of STC and scaled STC based on distortion-rate function. It is shown that for relatively small rates the ternarized sparsified codes almost achieve the Shannon distortion-rate function for i.i.d. Gaussian distributed data. In \cite{Sohrab_MLSTC_ICASSP2018_ArXiv}, the authors extended the concept of STC to multi-layer STC, a codebook-free scheme that successively refines the reconstruction of the residuals of previous layers. Based on the results in  \cite{Razeghi2018icassp} and \cite{Sohrab_MLSTC_ICASSP2018_ArXiv} we formulate our layered sparse coding that provides a multiple-access privacy-preserving identification scheme.
To this end, first we generate our first coebook with block-length $L$ and rate $R_1$ as: $\mathbf{U}_{(L, R_1)} \!=\! \varphi_1 \left( \mathbf{X} \right) \! = \!  \psi_{\lambda_1} \! \left( \mathbf{W}_1 \mathbf{X}\right)$. Next, we do reconstruction as: $\mathbf{\widehat{X}}^{[1]} = \varphi_{1}^{-1} \left( \mathbf{U}_{(L, R_1)} \right) = \mathbf{W}_1^{\dag} \mathbf{U}_{(L, R_1)}$. This provides the residual $\mathbf{R}^{[1]} \! = \!\mathbf{X}  - \mathbf{\widehat{X}}^{[1]}$. Now, we encode the residual of the first layer to generate the second codebook with block-length $L$ and rate $R_2$ as: $\mathbf{U}_{(L, R_2)} = \varphi_2 \left( \mathbf{R}^{[1]} \right) = \psi_{\lambda_2} \left( \mathbf{W}_{\!2} \mathbf{R}^{[1]}\right)$. The reconstructed data as well as the residual of the second layer obtained as  $\mathbf{\widehat{X}}^{[2]} = \varphi_{2}^{-1} \left( \mathbf{U}_{(L, R_2)} \right)$ and $\mathbf{R}^{[2]} = \mathbf{X} - \left( \mathbf{\widehat{X}}^{[1]} + \mathbf{\widehat{X}}^{[2]} \right)$, respectively. 
In the same way, the layered sparse coding scheme, initialized with $i =0$, can be formulated as:\vspace{-7pt}
\begin{eqnarray}
\mathbf{U}_{(L, R_{i+1})}  &= & \varphi_{i+1} \left( \mathbf{R}^{[i]} \right),\nonumber \\\vspace{-4pt}
\mathbf{\widehat{X}}^{[i]} &=& \varphi_{i}^{-1} \left( \mathbf{U}_{(L, R_i)} \right),  \nonumber \\\vspace{-4pt}
\mathbf{R}^{[i]} &=& \mathbf{X} - \sum_{j=1}^{i} \mathbf{\widehat{X}}^{[j]}. \vspace{-25pt}
\end{eqnarray}
Note that $\mathbf{\widehat{X}}^{[1]} \! \!   \rightarrow \!   \! \cdots  \!    \!  \rightarrow \!       \mathbf{\widehat{X}}^{[K]} $ forms a Markov chain. The algorithm successively refines the original database $\mathbf{X}$ over  (asymptotically large) $K$ stages, such that $\|  \mathbf{X} - \sum_{i=1}^K  \mathbf{\widehat{X}}^{[i]}\|_F^2 \leq \! N D$.


 \vspace{-5pt}
 
\subsection{Privacy Amplification Scheme}
 
 \vspace{-3pt}

The core idea of our privacy amplification scheme is to increase the general entropy of our sparse codes via adding some randomness to it. To this end, let $\mathcal{U}, \mathcal{V} \subseteq \mathcal{T}$ be two subspaces such that $\mathcal{T} = \mathcal{U} + \mathcal{V}$, where $\mathcal{T}$ is the space of $L$-dimensional sparse codes. So, every vector $\mathbf{u}_a  \in \mathcal{T}$ has at least one expression as $\mathbf{u}_a = \mathbf{u} + \mathbf{v}, \mathbf{u} \in \mathcal{U}, \mathbf{v} \in \mathcal{V}$. If we have $\mathcal{U} \cap \mathcal{V} = \left\{ \mathbf{0}  \right\}$, then every vector $\mathbf{u}_a \in \mathcal{T}$ has the \textit{unique} expression $\mathbf{u}_a = \mathbf{u} + \mathbf{v}, \mathbf{u} \in \mathcal{U}, \mathbf{v} \in \mathcal{V}$ and we write $\mathcal{T} = \mathcal{U} \oplus \mathcal{V}$. Also, $ \mathcal{T}$ is called the direct sum of $\mathcal{U}$ and $\mathcal{V}$. 
Now, let  $\mathcal{U}$ be the space of non-zero components of $\mathcal{T}$ and $\mathcal{V}$ be the space of zero components of $\mathcal{T}$. 
The idea of our ambiguization scheme is to set ambiguization noise $\mathbf{n}$ such that $\mathbf{n} \in  \mathcal{V}$. 
Furthermore, since $\left( \mathcal{T}, \langle  \cdot , \cdot \rangle \right)$ is an inner product space and $\mathcal{V} = \mathcal{U}^\bot \triangleq \left\{ \mathbf{v} \in \mathcal{V} : \langle \mathbf{u}, \mathbf{v} \rangle = 0, \forall \mathbf{u} \in \mathcal{U} \right\}$, $\mathcal{T} = \mathcal{U} \oplus \mathcal{U}^\bot$ is orthogonal direct sum of $ \mathcal{U}$ and $ \mathcal{U}^\bot$. It is clear that $\mathrm{dim}~\mathcal{T} = L$, $\mathrm{dim}~\mathcal{U} \!= \!S_x$ and $\mathrm{dim}~\mathcal{U}^\bot \!= \! L \! - \! S_x$. For more details about the performance of ambiguization scheme we refer the reader to \cite{Razeghi2017wifs}.

\vspace{-2pt} 
 
\subsubsection{Owner's Privacy Amplification}
 
Based on our definition, the data owner simply adds random samples with alphabet $\{ -1, +1 \}$ to the zero-components of his sparse codebook $\mathbf{U}_{(L,R)}$ and sends the ambiguized sparse codebook $\mathbf{U}_{(L,R_p)}$ to the public server. 
We denote by $S_{n_s}$ the sparsity level of ambiguization noise at the public server. Note that $0 \leq \! S_{n_s}  \! \leq L  \! - \! S_x$. 
Furthermore, the owner may send only a fraction $L_p < L$ of his sparse codes to the public server. In \cite{Razeghi2018icassp}, the authors analyzed this scheme with more details. In general, the public ambiguized sparse codebook generated as $\mathbf{U}_{(L_p, R_p)} \! =  \! A \left( \mathbf{U}_{(L, R)} \right)$ with the block-length $L_p$ and the rate $R_p$, where $A$ is an ambiguization function, which consists of randomness addition as well as codeword subspace selection. 

\vspace{-1pt} 
 
\subsubsection{Client's Privacy Amplification}
 
In order to prevent reconstruction of exact information about the client's interests at the public server side, the client ambiguizes his sparse code $\mathbf{b}$ by adding $S_{n_q}$ random samples with alphabet $\{ -1, +1 \}$ to the zero-components of his query. We denote by $\mathbf{b}_p = \mathbf{b} \oplus \mathbf{n}$ the public query. 
 
 \vspace{-5pt}

\begin{figure}[t]
    \centering
      \hspace{-8pt}
        \begin{subfigure}[h]{0.235\textwidth}
        \includegraphics[width=4.6cm, height=3.3cm]{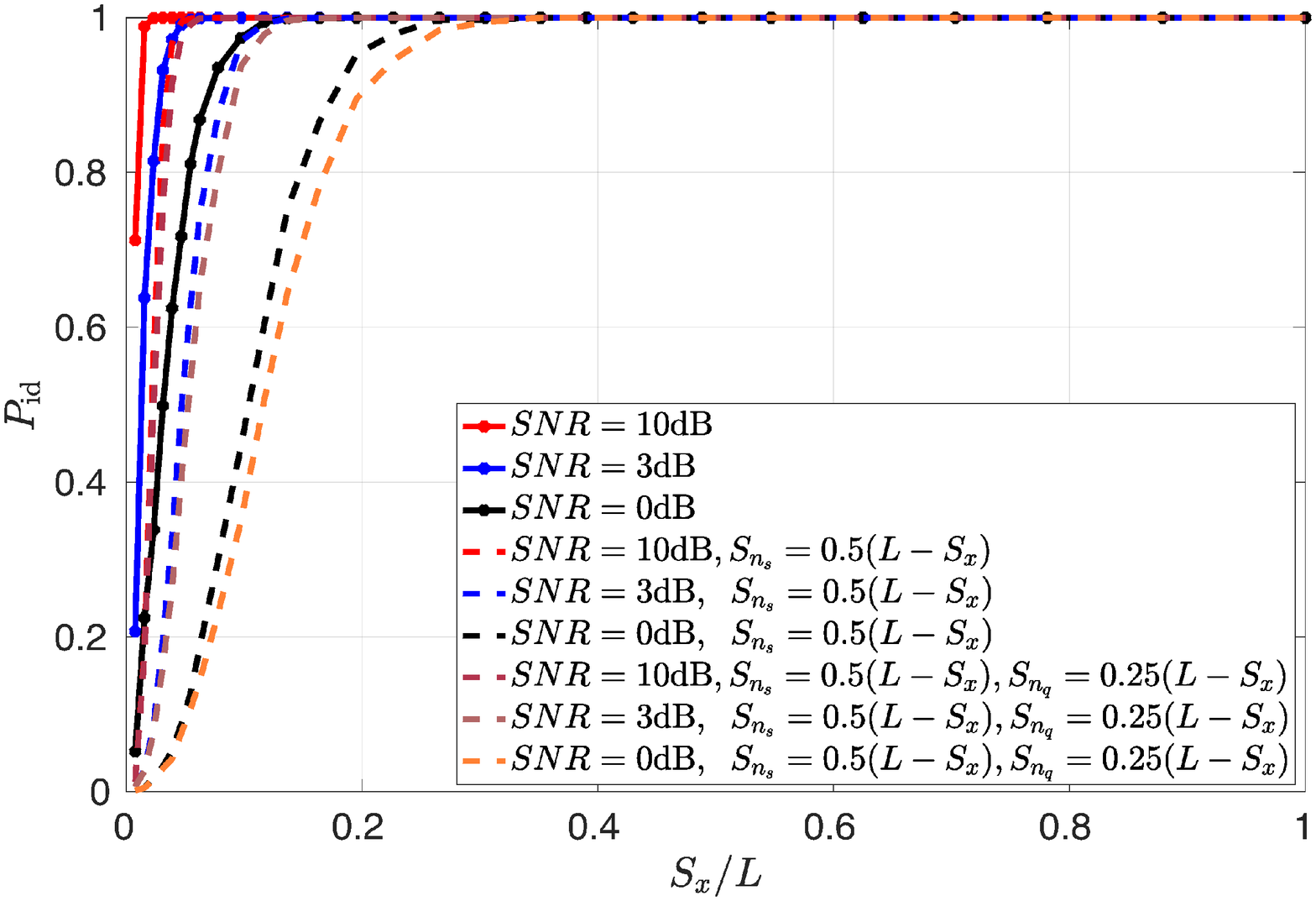}%
            \vspace{-3pt}
        \caption{ }
            \vspace{-5pt}
        \label{fig:p-IDvsSxPublic}
    \end{subfigure}%
  \hspace{-6pt}  ~   \hspace{-6pt}
       \begin{subfigure}[h]{0.235\textwidth}
        \includegraphics[width=4.6cm, height=3.3cm]{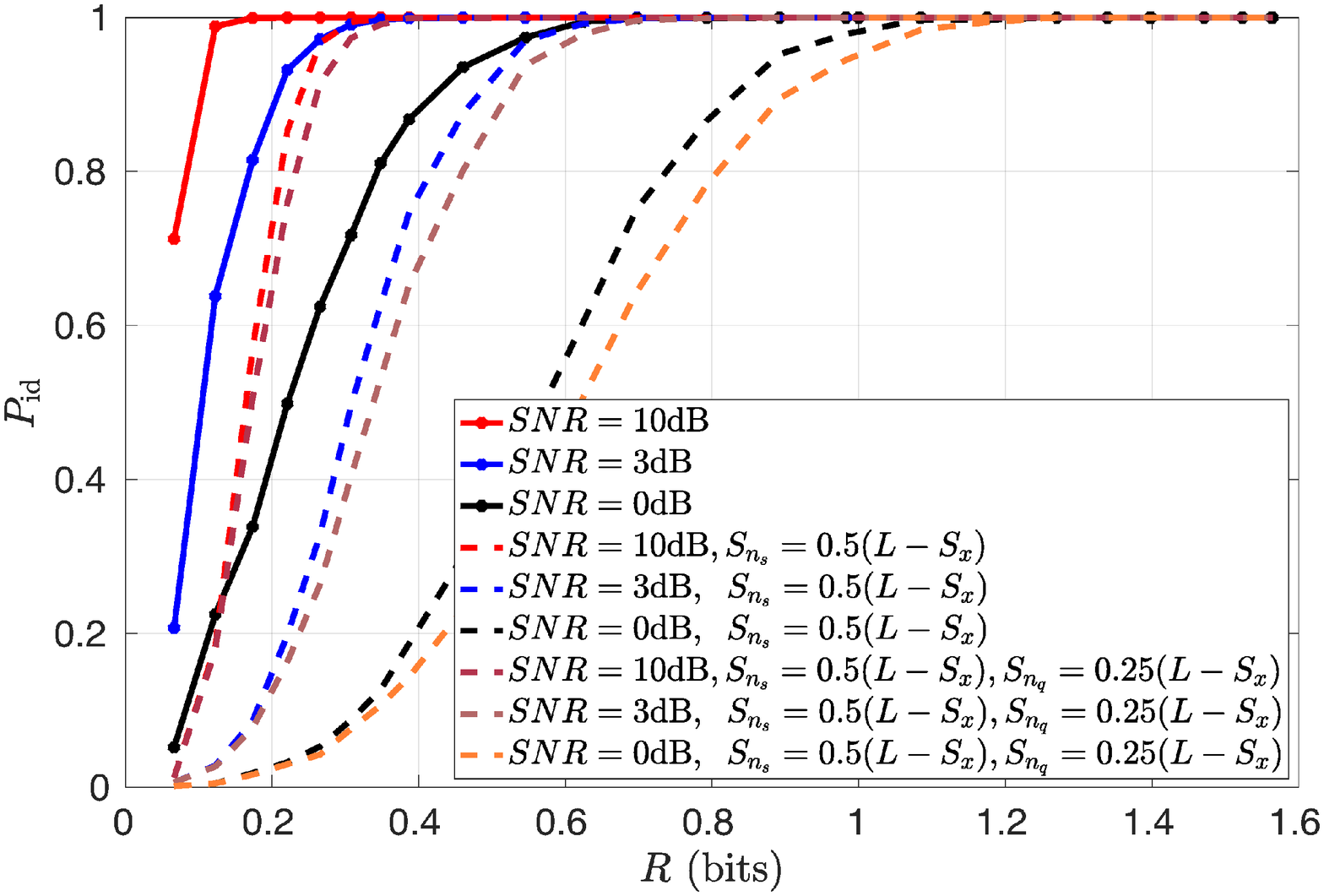}%
         \vspace{-3pt}
        \caption{ }
            \vspace{-5pt}
        \label{fig:p-IDvsRatePublic}
    \end{subfigure}
    \caption{The relation between probability of correct identification and a) sparsity ratio, b) encoding rate.}
    \vspace{-15pt}
    \label{Fig:P-correct-identificationPublic}
\end{figure}

\subsection{Algorithm}

 \vspace{-2pt}

 \subsubsection{Preparation at Owner Side}
\vspace{-2pt}

The owner generates offline the sparse codebook $\mathbf{U}_{(L, R_1)}$ with the \textit{trained linear} map $\mathbf{W}_1$ followed by the \textit{element-wise nonlinearity} thresholding operator $\psi_{\lambda_1} (\cdot)$, i.e., $\mathbf{U}_{(L, R_1)} = \psi_{\lambda_1} \left( \mathbf{W}_1 \mathbf{X} \right) = \varphi_1 \left( \mathbf{X} \right)$. Then, the owner performs the privacy amplification on codebook $\mathbf{U}_{(L, R_1)}$ to generate the public sparse codebook $\mathbf{U}_{a} = A \left( \mathbf{U}_{(L, R_1)} \right)$ with block-length $L_p$ and rate $R_p$. This ambiguized codebook is outsourced to the public server storage. 
Next, the owner generates successively the sparse codebooks $\mathbf{U}_{(L, R_i)}, i = 2, ..., K$ from the first sparse codebook $\mathbf{U}_{(L, R_1)}$. Therefore, the database $\mathbf{X}$ is encoded by total rate $R_1 + R_2 + \cdots + R_K$. 
The $K$ sparse codebooks $\mathbf{U}_{(L, R_i)}, i \in [K]$ are outsourced to the private server storage. The block diagram of codebooks generation is illustrated in Fig.~\ref{Fig:CodebookGeneration}.


 \vspace{-2pt}
 
 \subsubsection{Indexing at Server Sides}
The public and private servers index the received sparse codes. It can be indexed as in \cite{Sohrab_WIFS2016}. 

\vspace{-1pt}
 
\subsubsection{Querying at Client Side}
\vspace{-2pt} 
 
The client generates the sparse codeword $\mathbf{b}$ from its query $\mathbf{y}$, using the shared \textit{public trained linear} map $\mathbf{W}_1$ followed by the \textit{element-wise nonlinearity} operator $\psi_{\lambda_1}\left( \cdot \right)$, therefore $\mathbf{b} = \psi_{\lambda_1} \left( \mathbf{W}_1 \mathbf{y}\right)$. Then, the client ambiguizes his code by adding $S_{n_q}$ random samples with alphabet $\{ -1, +1 \}$ to his code. The ambiguized public query $\mathbf{b}_p = \mathbf{b} \oplus \mathbf{n}$ is send to the public server. The client also sends his original domain query $\mathbf{y}$ to the private server. Each client has a pre-defined authorization level at the private sever. 
 
 
\subsubsection{Initial Search at Public Server Side} 

\vspace{-2pt}

The public server seeks all $\{ \mathbf{u}_{a} \left(m\right) ,m \in \left[ M \right] \}$ ANNs in the radius $\gamma_p L_p$ from the query $\mathbf{b}_p$ in order to produce an initial public list $\mathcal{L}_p$ of possible candidates as $\mathcal{L}_p \left( \mathbf{b}_p \right) = \left\{ m \in \left[M\right] : d_{\mathcal{A}_p} \left( \mathbf{u}_a \left(m \right) , \mathbf{b}_p \right) \leq \gamma_p L_p \right\}$, 
where $ d_{\mathcal{A}_p} \left( \cdot , \cdot \right)$ is a similarity measure in space $\mathcal{A}_p$. Next, the public server sends the initial list $\mathcal{L}_p$ to the private server. 


\vspace{-2pt}

One can use different similarity measures $d_{\mathcal{A}_p} \left( \cdot , \cdot \right)$. However, due to many interesting properties, we consider a new similarity and dissimilarity measures based on the support intersection of the sparse codewords \cite{Sohrab_ISIT2017, kostadinov2018learning}.  To this end, we decompose sparse codes into positive part and negative part as $\mathbf{u} (m) = \mathbf{u}^{+} (m)  + \mathbf{u}^{-} (m)$ and $\mathbf{b}_{p} = \mathbf{b}_{p}^{+}  + \mathbf{b}_{p}^{-}$, where $\mathbf{u}^{+} (m) \! = \! \max \left(\mathbf{u}(m),  \mathbf{0}\right)$ and $\mathbf{b}_{p}^{+} \! = \! \max \left(\mathbf{b}_p,  \mathbf{0}\right)$ correspond to positive components and $\mathbf{u}^{-} (m)  \! =  \! \max \left(- \mathbf{u}(m),  \mathbf{0}\right)$ and $\mathbf{b}_{p}^{-} \! = \! \max \left(- \mathbf{b}_p,  \mathbf{0}\right)$ correspond to negative components. 
The similarity score $\mathrm{Sim}^{(m)}$ between $\mathbf{u} (m)$ and $\mathbf{b}_{p}$ is defined as:\vspace{-5pt}
\begin{equation}\label{Eq:SimMeasure}
\mathrm{Sim}^{(m)} = {\| \mathbf{u}^{+} (m) \odot  \mathbf{b}_{p}^{+}  \|}_1 + {\| \mathbf{u}^{-} (m) \odot  \mathbf{b}_{p}^{-} \|}_1, \vspace{-2pt} 
\end{equation}
and the dissimilarity score $\mathrm{Dis}^{(m)}$ between $\mathbf{u} (m)$ and $\mathbf{b}_{p}$ is defined as:\vspace{-5pt}
\begin{equation}\label{Eq:DisMeasure}
\mathrm{Dis}^{(m)} = {\| \mathbf{u}^{+} (m) \odot  \mathbf{b}_{p}^{-}  \|}_1 + {\| \mathbf{u}^{-} (m) \odot  \mathbf{b}_{p}^{+} \|}_1, \vspace{-2pt} 
\end{equation} 
where $\odot$ is the Hadamard product. For more details about the the theoretical aspects of the considered similarity measure, we refer the reader to \cite{kostadinov2018learning}. 
%
%
%

\vspace{-1pt}

The public list $\mathcal{L}_p$ is composed of the indices whose similarity score $\mathrm{Sim}^{(m)}, m \in [M]$ is higher than a threshold and dissimilarity score $\mathrm{Dis}^{(m)}, m \in [M]$ is below a threshold. 
Another option is to define a \textit{normalized similarity} as $\nu^{(m)} = \mathrm{Sim}^{(m)} /  ( \mathrm{Sim}^{(m)} +  \mathrm{Dis}^{(m)} ), \forall m \in \left[M \right]$. 
Therefore, the public list $\mathcal{L}_p$ is composed of the indices of the $\gamma$ largest $\nu^{(m)}$'s. 
Finally, the public server sends back the public list $\mathcal{L}_p$ to the private server. 
The public server can either fix the threshold or the number of $\gamma$ similar elements. 


\subsubsection{Multiple Access List Refinement at Private Server}
\vspace{-2pt}
 
The private server receives the public list $\mathcal{L}_p$, then it considers first layer codebook $\mathbf{U}_{(L, R_1)}$, which is the clean and full length version of the public codebook $\mathbf{U}_{a}$. Next, the private server reconstructs the codewords with indices reported on the public list as $\mathbf{\widehat{x}}^{[1]}(m) = \varphi_{1}^{-1} \left( \mathbf{u}_{(L,R_1)} (m)  \right), m \in \mathcal{L}_p$. 
It then computes the distance measure between private query $\mathbf{y}$ and reconstructed sparse codewords $\mathbf{\widehat{x}}^{[1]}(m), m \in \mathcal{L}_p$ in the original signal domain. This will produced the first private list $\mathcal{L}_{s_1} \left( \mathbf{y} \right) = \{ m \in \mathcal{L}_p: {\| \mathbf{y} - \mathbf{\widehat{x}}^{[1]} (m) \|}_2 \leq \gamma_{s_1} N \}$. Next, he reconstructs the codewords $\mathbf{u}_{(L, R_2)}(m), m \in \mathcal{L}_{s_1}$ of the second layer codebook $\mathbf{U}_{(L, R_2)}$. The second private list is obtained as $\mathcal{L}_{s_2} \left( \mathbf{y} \right) = \{ m \in \mathcal{L}_{s_1}: {\| \mathbf{y} - \left( \mathbf{\widehat{x}}^{[1]} (m) + \mathbf{\widehat{x}}^{[2]}\right) \|}_2 \leq \gamma_{s_2} N \}$. In the same approach, at the $K$-th layer the private list is given as $\mathcal{L}_{s_K} \left( \mathbf{y} \right) = \{ m \in \mathcal{L}_{s_{K-1}}: {\| \mathbf{y} - \sum_{i=1}^{K} \mathbf{\widehat{x}}^{[i]} (m)  \|}_2 \leq \gamma_{s_K} N \}$.

\begin{figure}[t]
    \centering
      \hspace{-8pt}
        \begin{subfigure}[h]{0.23\textwidth}
        \includegraphics[width=4.4cm, height=3.3cm]{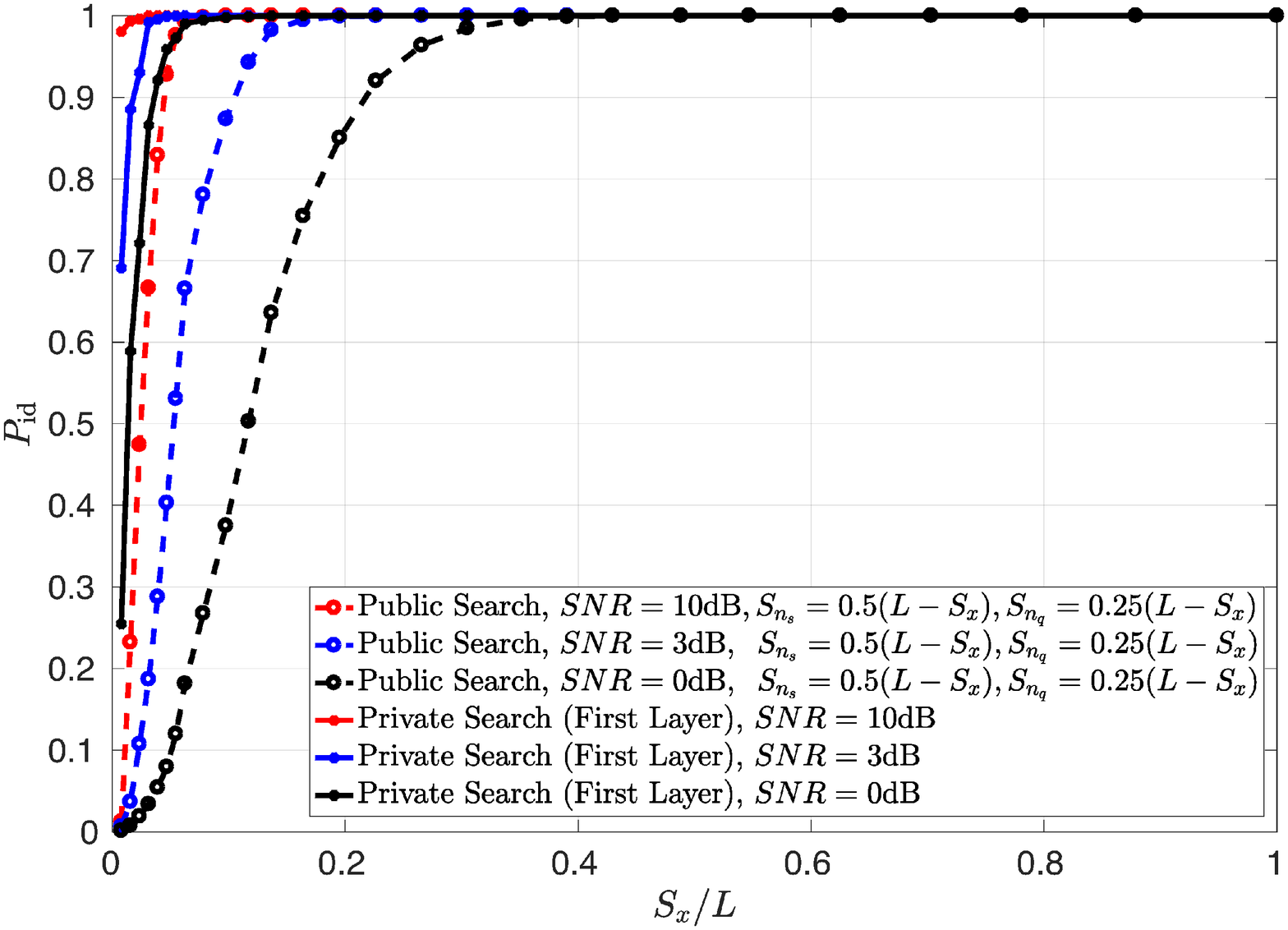}%
            \vspace{-3pt}
        \caption{ }
            \vspace{-5pt}
        \label{fig:p-IDvsSxPublicVsPrivate}
    \end{subfigure}%
  \hspace{-6pt}  ~   \hspace{-6pt}
       \begin{subfigure}[h]{0.23\textwidth}
        \includegraphics[width=4.4cm, height=3.3cm]{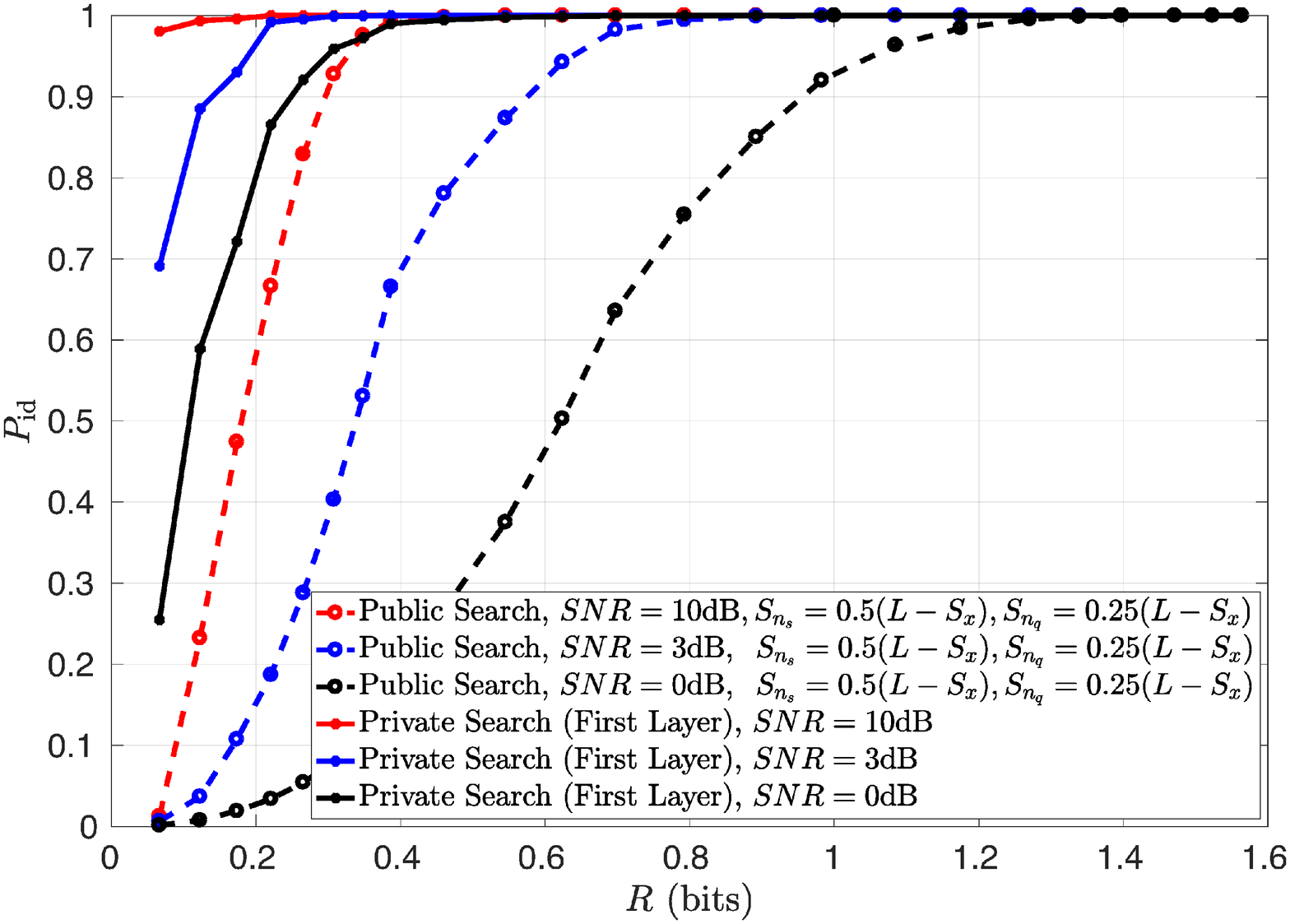}%
         \vspace{-3pt}
        \caption{ }
            \vspace{-5pt}
        \label{fig:p-IDvsRatePublicVsPrivate}
    \end{subfigure}
    \caption{Comparison between the probability of correct identification at the public server and private server.}
    \vspace{-15pt}
    \label{Fig:P-correct-identificationPublicVsPrivate}
\end{figure}

\vspace{-5pt}

\section{Performance Analysis}\label{Sec:IV}

\vspace{-2pt}

In this section we analyze the performance of our method in the terms of probability of correct identification $P_{\mathrm{id}}$ as well as privacy leakage. To this end we consider a database $\mathbf{X}$ of $M = 100\mathrm{K}$ random vectors with dimensionality $N=256$, which are generated from the distribution $\mathcal{N} \left( \mathbf{0},  \mathbf{I} \right)$. We then generate the noisy version of $\mathbf{X}$ with three different signal-to-noise-ratios (SNRs) $ 10 \mathrm{dB}$, $3 \mathrm{dB}$ and $0 \mathrm{dB}$, where $\mathrm{SNR} = 10 \log_{10} \frac{1}{\sigma_{\mathbf{z}}^2}$. We consider square sparsifying transform, i.e., $L=N$. Moreover, the sparsity level of the public sparse codewords as well as the public query code are considered the same.

\vspace{-1pt} 

In Fig.~\ref{Fig:P-correct-identificationPublic}, we depict the probability of correctly identifying the true query in the public list $\mathcal{L}_p$ as the function of sparsity ratio $S_x /L$ and encoding rate $R$. The red, blue and black solid lines show the performance of our method in the case that we impose no privacy amplification for the stored public database and the client's query. Next, we ambiguized our sparse public codebook by adding $S_{n_s} = 0.5 \left( L - S_x \right)$ random samples in the co-support of the public codewords. Finally, we complete our scenario by considering the privacy protection of query as well as owner's database., i.e., we ambiguized our codes by adding $S_{n_q} = 0.25 \left( L - S_x \right)$ samples in the co-support of the public query codewords.

\vspace{-2pt} 

In Fig.~\ref{Fig:P-correct-identificationPublicVsPrivate}, we compare the probability of correct identification at the public and private servers. We set the privacy amplification parameters of public codebook and client's query as $S_{n_s} \! =\! 0.5 \left( L - S_x \right)$ and $S_{n_q} \! =\! 0.25 \left( L \! - \! S_x \right)$, respectively. Then, we perform fast public search in the transform domain and send back the public list $\mathcal{L}_p$ to the private server. The results demonstrate high performance just by one layer similarity search in the original domain.

\begin{figure}[t]
    \centering
      \hspace{-8pt}
        \begin{subfigure}[h]{0.23\textwidth}
        \includegraphics[width=4.5cm, height=2.97cm]{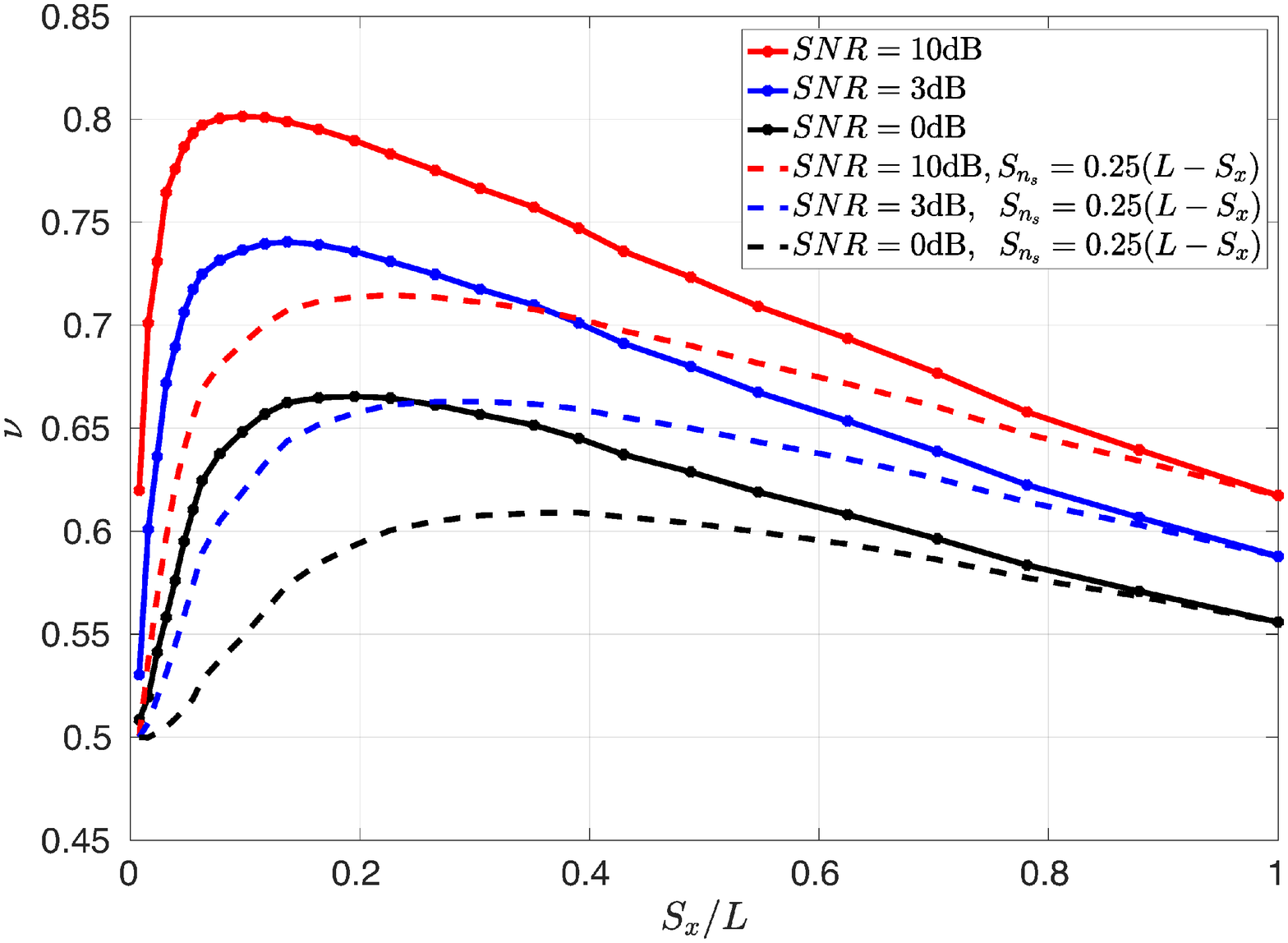}%
            \vspace{-5pt}
        \caption{ }
            \vspace{-5pt}
        \label{fig:relative-similarityVsSx}
    \end{subfigure}%
  \hspace{-6pt}  ~   \hspace{-6pt}
       \begin{subfigure}[h]{0.23\textwidth}
        \includegraphics[width=4.5cm, height=2.97cm]{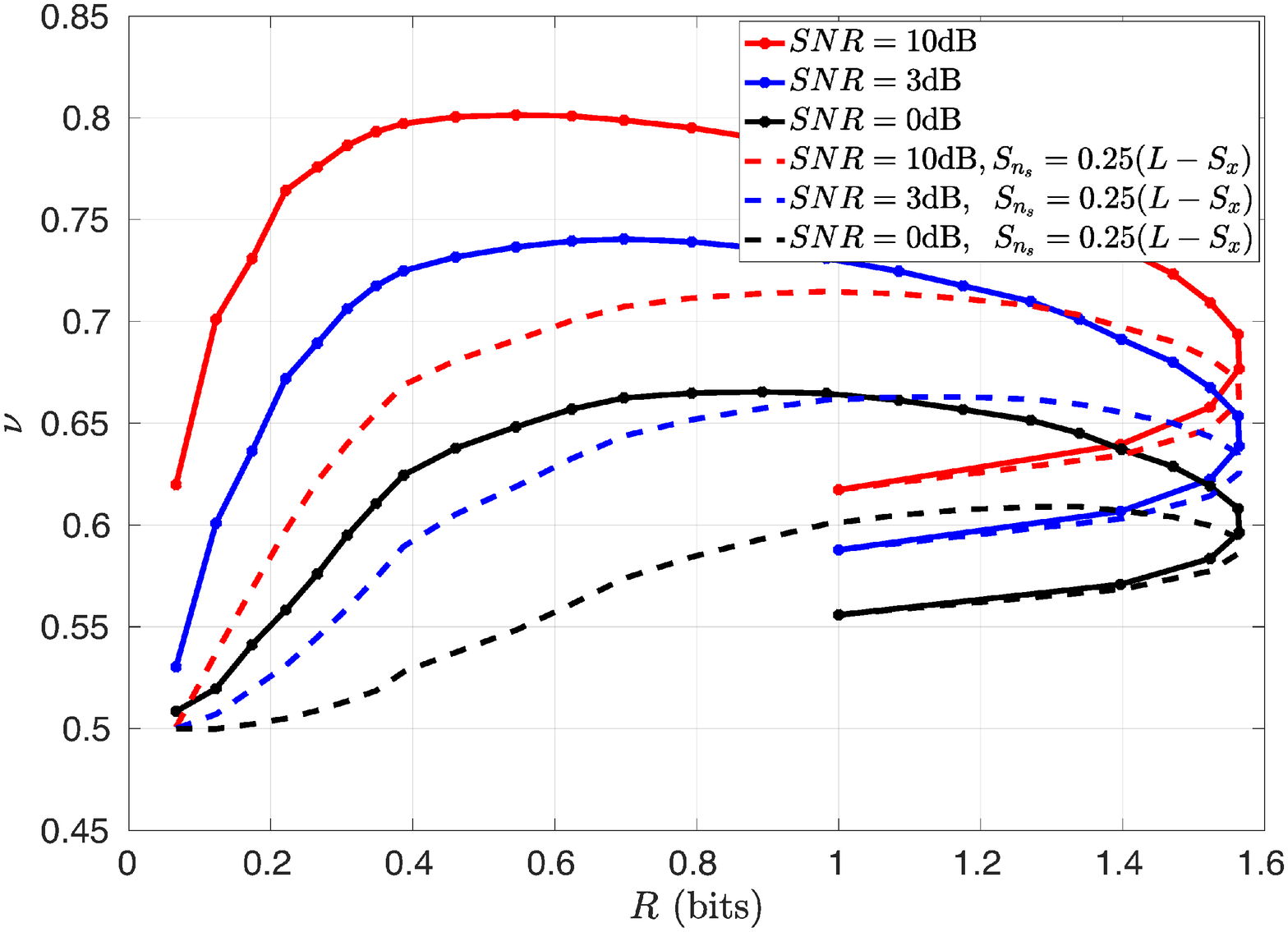}%
         \vspace{-5pt}
        \caption{ }
            \vspace{-5pt}
        \label{fig:relative-similarityVsRate}
    \end{subfigure}
    \caption{The relation between normalized similarity and: a) sparsity ratio, b) encoding rate.}
    \vspace{-15pt}
    \label{Fig:relative-similarity}
\end{figure}

\vspace{-2pt}

Based on the defined similarity measure in \eqref{Eq:SimMeasure} and \eqref{Eq:DisMeasure}, in Fig.~\ref{Fig:relative-similarity}, we illustrate the relation between \textit{normalized similarity} $\nu$ and sparsity ratio $S_x/L$ and encoding rate $R$. As it is shown, at the sparsity ratios (rates) close to zero the similarity measure grows much faster than dissimilarity measure such that we have the maximal normalized similarity for relatively small sparsity ratios (rates). However, after a certain level the dissimilarity measure grows faster than the similarity measure. 

\vspace{-2pt} 

\begin{figure}[t]
    \centering
      \hspace{-8pt}
        \begin{subfigure}[h]{0.23\textwidth}
        \includegraphics[width=4.5cm, height=3cm]{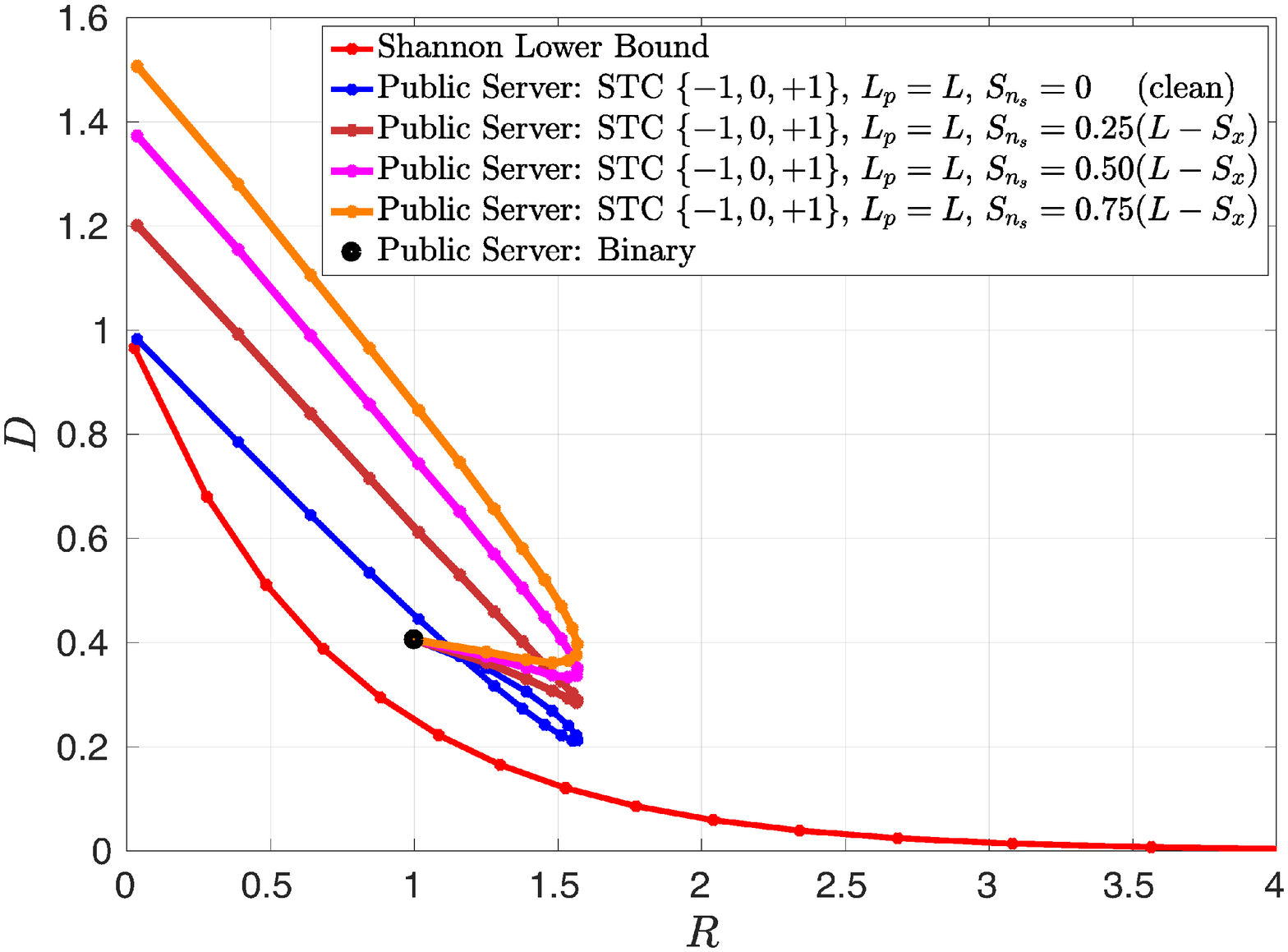}%
            \vspace{-4pt}
        \caption{ }
            \vspace{-6pt}
        \label{fig:distortion-ratePublic}
    \end{subfigure}%
  \hspace{-6pt}  ~   \hspace{-6pt}
       \begin{subfigure}[h]{0.23\textwidth}
        \includegraphics[width=4.5cm, height=3cm]{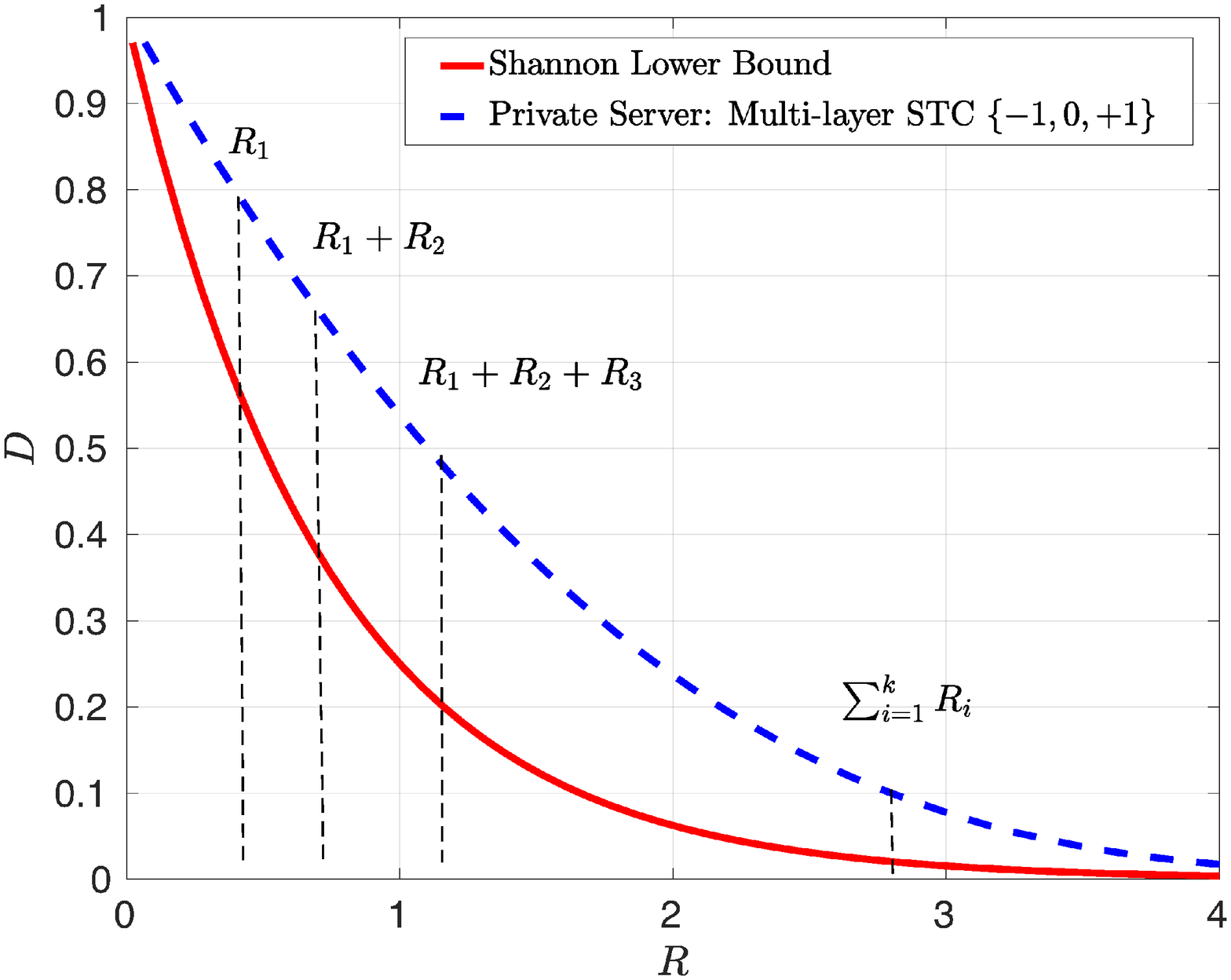}%
         \vspace{-4pt}
        \caption{ }
            \vspace{-6pt}
        \label{fig:distortion-ratePrivate}
    \end{subfigure}
    \caption{Distortion-rate behavior at the a) public server, b) private server.}
    \vspace{-15pt}
    \label{Fig:distortion-rate}
\end{figure}

In \cite{Razeghi2017wifs} and \cite{Razeghi2018icassp}, we defined the privacy measures in the terms of `reconstruction leakage' and `clustering leakage'. Based on the results in \cite{Razeghi2017wifs} and \cite{Razeghi2018icassp}, the curious public server cannot perform clustering the stored public database. Moreover, the un-authorized clients cannot infer the structure of database. 
%
In order to address the reconstruction leakage of the proposed privacy-preserving identification scheme, consider the mutual information between the random sequence $\mathbf{x}$, the reconstructed random sequence $\mathbf{\widehat{x}}_{a}$ at the public server and the reconstructed random sequences $\mathbf{\widehat{x}}^{[i]}, i \in [K]$ at the private server. Using the data-processing inequality and considering the markovity of random sequences, we have:\vspace{-5pt} 
\begin{eqnarray}
I  ( \mathbf{x}; \mathbf{\hat{x}}_a )  & \leq &  I  ( \mathbf{x}; \mathbf{\hat{x}}^{[1]}  )  \leq \cdots  \leq I  ( \mathbf{x}; \mathbf{\hat{x}}^{[1]}, \cdots  \mathbf{\hat{x}}^{[K]}  )\vspace{-8pt} . \nonumber 
\end{eqnarray}
In Fig.~\ref{Fig:distortion-rate}, we illustrate the distortion-rate behavior at the public and private servers, which interprets the `reconstruction' leakage in these scenarios. Fig.~\ref{fig:distortion-ratePublic} depicts the reconstruction leakage for three different ambiguization levels at the public sparse codebook and compares them with the Shannon lower bound. In Fig.~\ref{fig:distortion-ratePrivate}, we illustrate the performance of reconstruction at the private server and compare it with Shannon lower bound. This plot also depicts the accuracy of private lists for different authorization levels. Note that the illustrated results are obtained without considering any optimal rate allocation to our codebooks. By utilizing the optimal rate allocation and also multi-level quantization we can closely achieve the Shannon lower bound. That is beyond the scope of this paper.

\vspace{-5pt}

\section{Conclusion}\label{Sec:V}
 \vspace{-3pt} 
 
We have proposed a novel distributed privacy-preserving identification framework based on layered sparse codes with the ambiguization and granular access to the results of identification. The initial fast search is performed on the public server and the refined searches are performed on the distributed private server(s). The accuracy of the private search is based on the authorization level of the clients. The results show the performance of proposed scheme in the terms of probability of correct identification as well as the privacy leak measures. 


%


 \vspace{-4pt}

\bibliographystyle{IEEEtran}
\bibliography{references}
%
%
%

\end{document}